\documentclass[conference]{IEEEtran}
\IEEEoverridecommandlockouts
\usepackage{cite}
\usepackage{amsmath,amssymb,amsfonts}
\usepackage{algorithmic}
\usepackage{graphicx}
\usepackage{textcomp}
\usepackage{xcolor}
\usepackage{listings}
\usepackage[disable]{todonotes}
\newcommand{\comment}[1]{}

\def\BibTeX{{\rm B\kern-.05em{\sc i\kern-.025em b}\kern-.08em
    T\kern-.1667em\lower.7ex\hbox{E}\kern-.125emX}}
    
\makeatletter
\def\endthebibliography{%
  \def\@noitemerr{\@latex@warning{Empty `thebibliography' environment}}%
  \endlist
}
\makeatother
    
\begin{document}

\title{Optimising AI Training Deployments using Graph Compilers and Containers}

\author{
\IEEEauthorblockN{Nina Mujkanovic }
\IEEEauthorblockA{\textit{HPE HPC/AI EMEA Research Lab} \\
Basel, Switzerland \\
nina.mujkanovic@hpe.com}
\and
\IEEEauthorblockN{Karthee Sivalingam}
\IEEEauthorblockA{\textit{HPE HPC/AI EMEA Research Lab} \\
Bristol, United Kingdom \\
karthee.sivalingam@hpe.com}
\and
\IEEEauthorblockN{Alfio Lazzaro}
\IEEEauthorblockA{\textit{HPE HPC/AI EMEA Research Lab} \\
Basel, Switzerland \\
alfio.lazzaro@hpe.com}
}

\maketitle

\begin{abstract}

Artificial Intelligence (AI) applications based on Deep Neural Networks (DNN) or Deep Learning (DL) have become popular due to their success in solving problems like image analysis and speech recognition. Training a DNN is computationally intensive and High Performance Computing (HPC) has been a key driver in AI growth. Virtualisation and container technology have led to the convergence of cloud and HPC infrastructure. These infrastructures with diverse hardware increase the complexity of deploying and optimising AI training workloads. AI training deployments in HPC or cloud can be optimised with target-specific libraries, graph compilers, and by improving data movement or IO. Graph compilers aim to optimise the execution of a DNN graph by generating an optimised code for a target hardware/backend.

As part of SODALITE (a Horizon 2020 project), MODAK tool is developed to optimise application deployment in software defined infrastructures.  Using input from the data scientist and performance modelling, MODAK maps optimal application parameters to a target infrastructure and builds an optimised container. In this paper, we introduce MODAK and review container technologies and graph compilers for AI. We illustrate optimisation of AI training deployments using graph compilers and Singularity containers. Evaluation using MNIST-CNN and ResNet50 training workloads shows that custom built optimised containers outperform the official images from DockerHub. We also found that the performance of graph compilers depends on the target hardware and the complexity of the neural network. 
\end{abstract}


\section{Introduction}
Increasing availability of data and the computational power of High Performance Computing (HPC) have driven the adoption of Artificial Intelligence (AI) in recent years. 
Deep Learning (DL), a subset of AI that uses multi-layers of neural networks to progressively extract higher level features from raw data, has dramatically improved the state-of-the-art in domains like speech recognition, visual object recognition, and many 
others\cite{alom2018history,lecun2015deep}.  
This growth started with the success of the Convolutional Neural Network (CNN)\cite{fukushima1980neocognitron} \emph{AlexNet}\cite{krizhevsky2012imagenet} in 2012.
\emph{AlexNet} is computationally expensive and used GPUs to accelerate the training time. Frameworks like  \emph{TensorFlow}\cite{abadi2016tensorflow}, \emph{PyTorch}\cite{NEURIPS2019_9015},  \emph{MXNet}\cite{chen2015mxnet}, and \emph{CNTK}\cite{seide2016cntk} simplify the development and deployment of DL training workloads. Some frameworks also support a high-level language like \emph{Keras}\cite{chollet2015keras}.

In recent years, DL training networks have grown in size and complexity. With exponential increase in data and use cases, AI training workloads are being deployed across heterogeneous hardware targets like HPC and cloud. The user experiences in cloud environments and on HPC systems differ vastly. Cloud environments offer a number of Domain Specific Language (DSL) based tools such as \emph{Terraform}\cite{terraform} and \emph{Cloudify}\cite{cloudify_2020} to simplify the management of the entire application life-cycle, including the deployment, monitoring, and maintenance of application models. HPC systems on the other hand require specialist knowledge of the system and command line tools to manage the application lifecyle. They are accessed using tools such as Secure Shell (SSH), and require job submission to compute nodes via workload managers like SLURM~\cite{Jette02slurm:simple} and TORQUE~\cite{klusavcek2015planning}. This can be a high hurdle for domain scientists interested in running experiments in HPC systems compared to cloud. 

The cross-section of developing, optimising, and deploying AI applications across heterogeneous infrastructures like HPC or cloud environments poses a complex problem.  EU projects of the Heterogeneity Alliance\cite{hetero} like COLA\cite{cola}, TANGO\cite{tango}, HiDALGO\cite{hidalgo}, Exa2Pro\cite{exa2pro}, EcoScale\cite{ecoscale} and SODALITE aim to deliver software tools, methods, and knowledge to enable next-generation applications to use heterogeneous hardware. SODALITE\cite{sodalitewebsite}, a European Horizon 2020 project, aims to solve the problem of deploying workloads across heterogeneous environments by providing tools for software-defined infrastructures.
In SODALITE, we have developed MODAK, a model-based application deployment optimiser for static optimisation in software defined infrastructure. 
In this paper we introduce MODAK and evaluate its usage for optimising AI training workloads using graph compilers and containers.

\begin{figure*}[hbt!]
    \centering
    \includegraphics[width=0.9\textwidth]{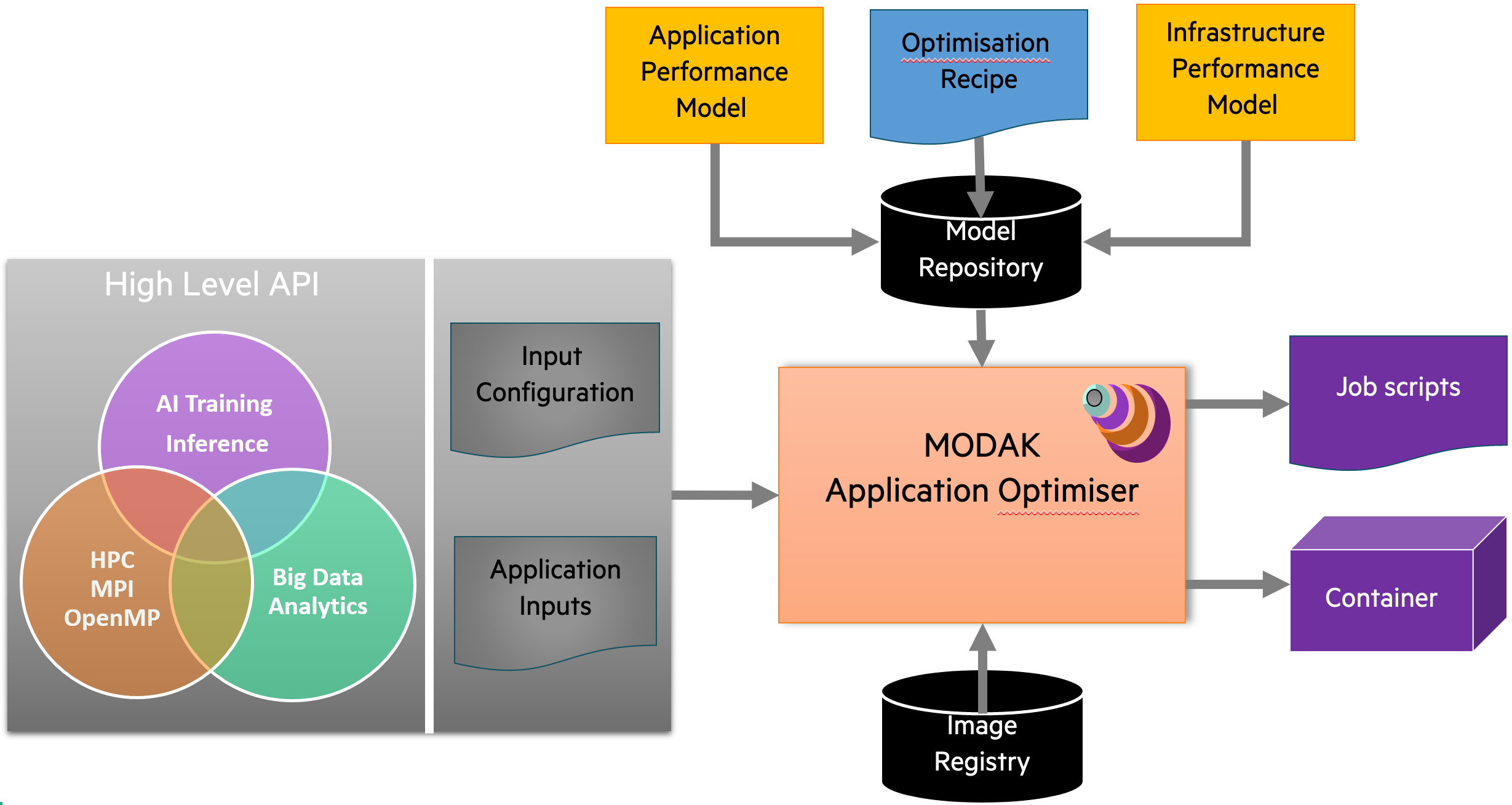}
    \caption{MODAK architecture}
    \label{fig:modak_arch}
\end{figure*}

\section{Related Work}
Application performance and scalability are important for HPC users. The optimisation process generally involves manual profiling and tuning of application parameters to suit target hardware. Furthermore, it is not portable and needs to be repeated when moving to other HPC systems due to the diversity of hardware in HPC systems. 

The automation of application optimisation on both HPC and cloud systems requires models that can be used for performance prediction and to study how different hardware components affect performance, a task made more complex by the wide variety of cloud offerings available with a wide variety of hardware. Application profiling and historical data gathered on HPC and cloud systems were used by \cite{baughman2018profiling} to 
create a performance model. \emph{ParaOpt}\cite{wu2019paraopt}, a tool that autotunes application configurations for different instance types based on runtime and cost, was evaluated for genomics, molecular dynamics, and machine learning applications on multiple public clouds. 

A number of works explored the performance of cloud environments. Exabyte compared cloud targets using the Linpack benchmark\cite{mohammadi2018comparative}, and developed a software tool for the continuous evaluation of various cloud environments\cite{mohammadi2018continuous}. EPCC directly compared the performance of HPC on-premise systems and the Oracle cloud cluster using the DiRAC application benchmarks~\cite{epcc}, discovering issues in the usability and scalability of cloud based clusters. 

Many tools are developed to optimise application deployments that are packaged as containers. \emph{ConfAdvisor}\cite{chiba2019confadvisor} is a tuning framework for containers on Kubernetes. AWS compute optimiser\cite{awscompute} optimises workloads for both cost and performance based on historical utilization metrics. 
Google~\cite{krishnan2015google} similarly offers optimised containers for AI application deployments on the Google Cloud Platform.
\emph{HPAI} project~\cite{brayford2019deploying} studied the feasibility of deploying AI workloads in HPC systems using \emph{Charliecloud}\cite{priedhorsky2017charliecloud}. 

In the next section, we will introduce MODAK, an application optimiser for software defined infrastructures like HPC and cloud.

\section{MODAK}
In a software defined world, where software rules and hardware is abstracted, enabling applications to optimally run on diverse targets gives flexibility and saves money and time. SODALITE\cite{sodalitewebsite}, a European Horizon 2020 project, aims to solve the problem of deploying workloads across heterogeneous environments by providing tools for software-defined infrastructures (SDI) that place the computing infrastructure under software control, abstracting away hardware dependencies. This simplifies and improves user exploitation of heterogeneous targets like HPC clusters, cloud environments, and edge devices.

MODAK is the SODALITE component responsible for enabling static optimisation of application deployment in a software defined way. The performance of an application when deployed in a specific infrastructure can be predicted using performance models of the application and infrastructure. The performance models are developed by running standard benchmarks across different configurations of both the application workload and the deployment infrastructure, and then building a linear statistical model. This model informs MODAK about how the application parameters, such as the input data size and format, affect the performance relative to the performance characteristics of the target infrastructure, such as peak performance and memory bandwidth. Using this knowledge, MODAK maps the optimal application parameters to the infrastructure target and builds an optimised container. 

Figure~\ref{fig:modak_arch}  shows the MODAK architecture and its dependencies. Application performance optimisation is highly dependent on the application, its configuration, and the infrastructure. MODAK supports three major application types for static optimisation - AI training and inference, Big Data Analytics, and traditional HPC. The data scientist, or AoE, selects application optimisations using the SODALITE IDE. Optimisations include changes to the application configuration, the environment, or the runtime. Application runtime parameters can be further autotuned for improved application performance. 

In order to apply the optimisations, the Application Optimiser requires that application code be written in a standard high-level API, along with the application inputs and configuration. This enables the Optimiser to make performance decisions based on the available target. The Optimiser uses the pre-built, optimised containers from the Image Registry and modifies them to build an optimised container for the application deployment. The Application optimiser also makes changes to runtime, deployment, and job scripts for submission to HPC schedulers. Describing all the components of MODAK is out of scope of this paper and a detailed description can be found in~\cite{modak}.

\section{Background} 
This subsection provides a brief contextual overview of the container solutions, AI frameworks, and graph compilers that are modelled for AI training optimisation for MODAK. 

\subsection{Container technology}
Containers are a technology with roots in the Unix chroot command released in 1979\cite{bernstein2014containers}. Linux containers (LXC), a precursor to the below listed technologies, use OS-level virtualization to isolate processes and resources in separate user namespaces\cite{helsley2009lxc}, a virtualization technique with a far lower overhead and higher scalability than hypervisor virtualization\cite{joy2015performance}. 

\emph{Docker}\cite{merkel2014docker} performs virtualization using LXC for kernel-level namespace isolation and cgroups for resource control. While it is an industry standard in cloud environments, its design poses security and performance issues on HPC systems. In particular, it does not support multi-user HPC systems, and its use of root daemons to build and run containers enables users to gain privileged access to the host systems network filesystem. 

\emph{Shifter}\cite{gerhardt2017shifter} and \emph{Charliecloud}\cite{priedhorsky2017charliecloud}, developed at NERSC and LANL respectively, were both designed with HPC systems in mind, making them more suitable to traditional HPC workflows. \emph{Charliecloud}, in particular, is a lightweight containerization technology based on a User Defined Software Stack (UDSS) and developed around the strict security requirements posed by sites. Drawbacks include a high administrative overhead (\emph{Shifter}) and a current lack of community uptake (\emph{Charliecloud}).

\emph{Singularity}\cite{sylabswebsite}\cite{kurtzer2017singularity}, developed by Berkeley National Laboratories, appears a good compromise between the cloud standard \emph{Docker} and the HPC-specific \emph{Shifter} and \emph{Charliecloud}. Built for HPC systems, and offering native support for HPC components including resource managers (\emph{Slurm}, \emph{Torque}, etc.), job schedulers, and some MPI features, it also offers an easy containerization workflow for users. Its privilege model relies on SUID and non-privileged user namespaces to launch containers as child processes, thus allowing for non-root users to create and launch containers safely. \emph{Singularity} can import and run Docker images directly.

We chose \emph{Singularity} as the optimal solution container deployment, with plans to extend to \emph{Docker}. Further reading on container technologies can be found in \cite{rudyy2019containers,benedicic2019sarus}.

\subsection{Graph compilers}

An approach all AI frameworks have in common is the use of intermediate representations (IR) to represent the neural network models as computational graphs, with nodes representing tensor operations and edges the data dependencies between them. There are typically multiple levels of IR, with high-level IRs residing in the frameworks' user-facing front-end, and low-level IRs residing in the back-end. 

A set of framework specific compilers can be used to perform optimisations on the generated graph IRs. These graph compilers can be grouped into two types - low-level tensor compilers, focused on the construction of high-performance operators for compute intensive operations, and deep learning compilers, focused on high-level optimisations on the IR followed by offloading to vendor specific libraries.

\emph{XLA} (Accelerated Linear Algebra)\cite{TFXLAwebsite}\cite{leary2017xla} is a TensorFlow specific graph compiler that accelerates linear algebra. It accepts a graph defined in the High Level Optimiser (HLO) IR and performs target-independent optimisation and analysis on it, such as operation fusion and buffer analysis. The optimised HLO IR is then sent to the back-end, which performs further HLO-level optimisations targeted to the hardware. The final code is generated using LLVM.

\emph{GLOW}\cite{rotem2018glow}, short for graph lowering, optimises PyTorch models by lowering the graph into a two-phase IR, with strong focus on the low-level IR. The high-level IR is then used for domain-specific optimisations, while the low-level instruction-based, address-only IR is used to perform memory-related optimisations. Machine specific code is generated at the lowest level. 

\emph{nGraph}\cite{cyphers2018intel} is a framework independent graph compiler that can be used by TensorFlow, PyTorch, and a number of other frameworks. It acts as a bridge, porting the framework specific graph model to a common, intermediate high-level IR that is then used to generate code optimised for a specific back-end via vendor-specific libraries.

\begin{figure*}[ht]
    \centering
    \includegraphics[width=0.8\textwidth]{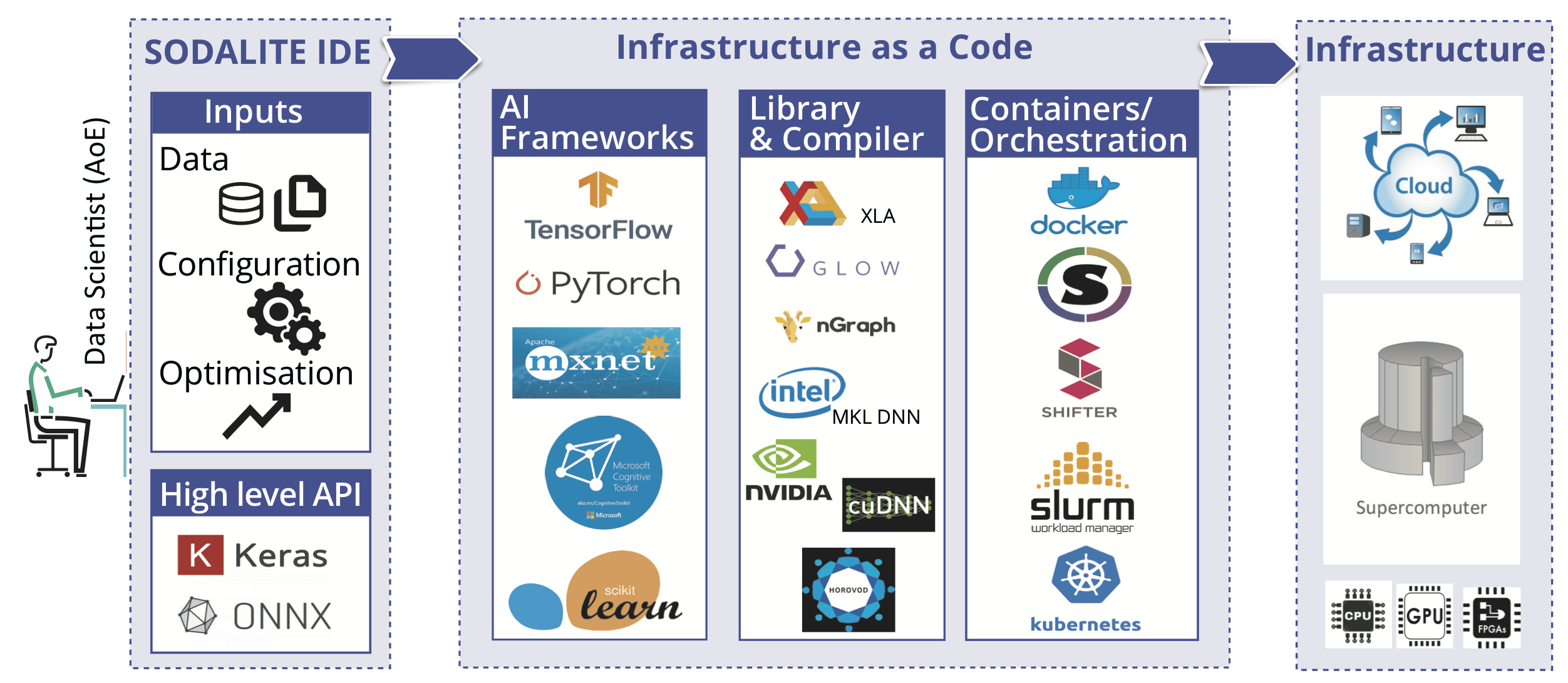}
    \caption{SODALITE AI example}
    \label{fig:sodalite_ai}
\end{figure*}

\section{Methodology}
In this section, we demonstrate the usage of MODAK for optimising AI training workloads using containers and graph compilers.
\subsection{MODAK AI Training Example}
Figure~\ref{fig:sodalite_ai} shows an example usage of MODAK for AI training application deployment. 
A data scientist uses the SODALITE IDE to build the application model with input data, configuration, and optimisations. The application, written in a high-level language or API like Keras or \emph{ONNX}\cite{lin2019onnc}, is then deployed in a \emph{Docker} or \emph{Singularity} container with the selected AI framework, using optimised libraries like \emph{MKL}\cite{wang2014intel} or \emph{cuDNN}\cite{brown2015gpu}, and compilers like \emph{XLA} and \emph{GLOW}. 
This optimised, containerised application is then deployed to an HPC or cloud system. 

In the IDE, the data scientist encodes optimisation options in a \emph{json} format, which is input to MODAK. Listing~\ref{lst:opt_dsl} shows a section of an optimisation DSL for a TensorFlow deployment. Here, custom build optimisations for a selected target (\emph{x86} and \emph{NVIDIA}) and the \emph{XLA} compiler are enabled. 
MODAK prebuilds TensorFlow containers and tags them based on supported optimisations.
Based on the selected optimisations in the DSL, MODAK selects the optimised container. MODAK can also build a container during deployment. 

\begin{lstlisting}[caption={Example Optimisation DSL},label={lst:opt_dsl}]

"optimisation": {
    "enable_opt_build": true,
    "app_type": "ai_training",
    "opt_build": {
        "cpu_type": "x86",
        "acc_type": "Nvidia"},
    "ai_training": {
      "tensorflow": {
        "version": "1.1",
        "xla": true }}}
\end{lstlisting}

\subsection{Containers for AI frameworks}
For MODAK, we created and evaluated AI framework containers on the SODALITE HPC testbed set up at \emph{HLRS}, the research and supercomputing center affiliated to the University of Stuttgart\cite{hlrs}. The testbed consists of a front-end node running Torque, and five compute nodes, each hosting an Nvidia GeForce GTX 1080 Ti GPU, an Intel(R) Xeon(R) CPU E5-2630 v4 processor, and 125GB of main memory. All nodes allow access via https. 

The container runtime installed on the system is Singularity version 3.4.1-1. In order to build and run containers, an administrator has to add additional mappings to the \emph{Singularity} user namespace UID and GID files to enable the use of \verb|fakeroot|, a feature used to impersonate root without superuser escalation\cite{singularitymapping}.

We determined which frameworks and graph compilers to benchmark based on popularity, availability of images, clarity of build instructions and documentation. Table~\ref{tab:ai-src} lists the set of AI framework images, graph compilers, versions, and their source. Official project images were downloaded from DockerHub (Hub) or packaged as Singularity containers using the Python package manager (\emph{pip}) or by following the source build instructions available on the project websites (opt-build). We ensured that optimisation libraries were available and matched across the official and our images (e.g. TensorFlow \emph{XLA}, etc.). Note that, while \emph{XLA} is listed separately in the table, it is auto-built as part of the TensorFlow framework. For \emph{XLA} and \emph{nGraph}, the supported TensorFlow version is specified. 
\begin{table}
\caption{\label{tab:ai-src}Source of AI Framework containers.}
\begin{center}
\begin{tabular}{|p{2cm}|p{1cm}|p{1cm}p{1cm}p{1.2cm}|}
 \hline
 AI Framework & version & Hub & pip & opt-build\\
 \hline\hline
 TensorFlow &1.14   &     &X&  X \\
 TensorFlow &2.1 &   X  & X   & X\\
 PyTorch &1.4 & X & X &  X\\
 MXNet &1.6& X & & \\
 CNTK &2.7& X & & \\
 XLA &2.1& X & X &  X\\
 GLOW &NA&      &  & X\\
 nGraph &1.14&    & X   &\\
 \hline
\end{tabular}
\end{center}
\end{table}

The DockerHub containers were retrieved using the \emph{Singularity} pull command, which directly ports \emph{Docker} containers to Singularity \verb|.sif| files. We chose the images tagged with the required \verb|version| and a \emph{cpu} or \emph{gpu} \verb|target| to establish a baseline for comparison.

To build the remaining containers, we created Singularity definition files. The definition files are composed of a header that describes the operating system (OS) used within the container, and multiple sections for pre-build setup, file importation into container, container environment setup, post OS installation container commands, etc.

As dependencies differ depending on whether containers execute CPU or GPU workloads, we developed two base OS containers to be called on in the definition header and used by our custom built containers.

\subsection{Containers for CPU}

The CPU enabled containers use a custom built Ubuntu 18.04 image as the base OS in the header. The image includes the packages \emph{llvm-8}, \emph{clang-8}, and \emph{Python3}. The remaining build instructions are encoded in the post OS installation section. For the pip based containers, this includes adding commands that install additional dependencies, followed by the \emph{pip} command to install the framework or compiler as per project instructions. 

Similarly, the source build containers have all installation instructions encoded in the post section and adhere to the instructions given by the individual projects' documentation\cite{tfsourcebuild,torchsourcebuild,glowsourcebuild}. Where applicable, compiler optimisation flags were set to improve performance on the CPU. TensorFlow, specifically, uses the build tool \emph{Bazel}, which accepts compiler flags via the argument \verb|--copt|.

\subsection{Containers for GPU}
The GPU containers use NVIDIA DockerHub images containing the \emph{NVIDIA-kernel}, \emph{cuda toolkit 10.1}, \emph{cudNN7}, and Ubuntu 18.04 as the base OS. We chose the NVIDIA base image to avoid portability issues and ease dissemination, as it is not possible to retrieve \emph{cudNN7} via the command line. All NVIDIA package paths are then set in the container environment section to enable their use for source builds. 

The remainder of the container build file is similar to that of the CPU containers. The source build commands are set in the post section of the file, with changes made to reflect differences in GPU builds, especially pertaining to the TensorFlow build. 

All containers - CPU and GPU - are then built using the \emph{Singularity} build command with the \verb|--fakeroot| flag set. Depending on the framework or graph compiler, this build can take from a couple of minutes to multiple hours. The built containers can then be run interactively, or launched to execute specific commands. Note that \emph{Singularity} places strong restrictions on GPU containers - the container must have the nvidia-kernel and any other dependencies such as \emph{CUDA} installed, and the nvidia-kernel version must match the host nvidia-kernel version. This requirement can be circumvented by using the \emph{Singularity} NVIDIA flag \verb|--nv| when launching containers. 

\subsection{Benchmarks}
We measured container performance by performing image classification training and timing the execution of a set number of epochs. To properly assess the frameworks on both CPU and GPU, we chose to train on two datasets, \emph{MNIST}\cite{lecun-mnisthandwrittendigit-2010} for the CPU workload, and \emph{ImageNet}\cite{deng2009imagenet} for the GPU workload. 

\emph{MNIST} training is an image classification problem for handwritten digits \cite{deng2012mnist}\cite{lecun1998gradient}. \emph{MNIST} itself refers to a dataset of 60,000 gray-scale images containing handwritten digits from 0 to 9.

\emph{ImageNet} is a database that consists of more than 14 million hand-annotated images in 20,000 categories. The dataset, published in 2009, eventually evolved into the annual ImageNet Large Scale Visual Recognition Competition (\emph{ILSVRC})\cite{russakovsky2015imagenet}, at which \emph{AlexNet}\cite{krizhevsky2012imagenet} achieved a novel 15.3\% top-5 error rate in 2012, setting off a longterm trend of running DL workloads on GPUs. 

The \emph{MNIST} dataset was trained on a CNN consisting of a combination of two convolutional layers, two \emph{maxpool} layers, two fully connected layers, and a \emph{softmax} activation function. For all benchmarks, we used a batch size of 128, image size of (28,28), and trained the network with 1,199,882 trainable parameters for 12 epochs.

\begin{figure}[htb!]
    \centering
    \includegraphics[width=6 cm]{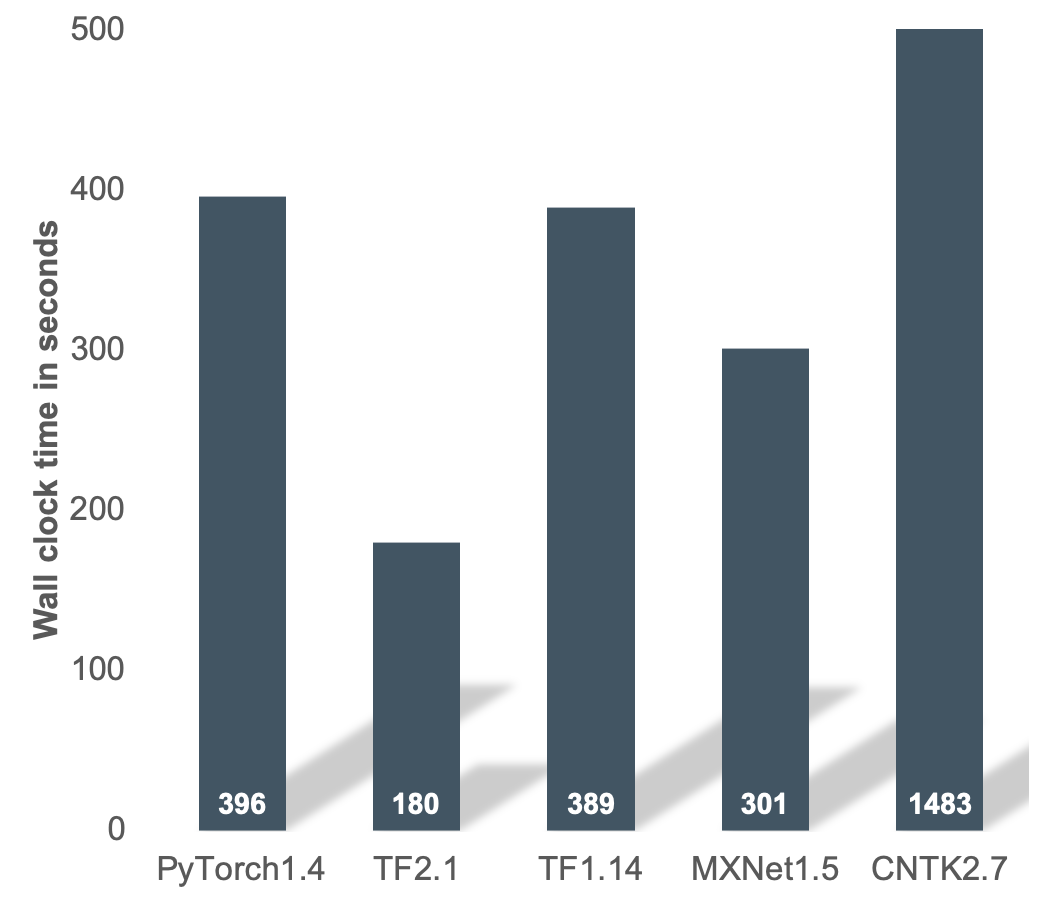}
    \caption{Performance of various AI framework containers on CPU MNIST training workload}
    \label{fig:mnist_all}
\end{figure}

For the \emph{ImageNet} training, we used the \emph{ResNet50}\cite{he2016deep} residual network, a fifty-layer deep neural network. Residual networks use skip connections to pass residual functions, minimizing the problem of vanishing or exploding gradients\cite{hochreiter2001gradient} and permitting the use of far deeper networks. For all \emph{ResNet50} benchmarks, we used single precision, a batch size of 96, and trained for 3 epochs.

We wrote the \emph{MNIST} and \emph{ImageNet} workloads using the DSL of each framework and graph compiler. The workloads were submitted to one node exclusively per job using a \emph{Torque} submission file. The training was not performed to convergence but instead stopped after 12 epochs for \emph{MNIST}, and 3 epoch for \emph{ImageNet}. This considerably shortened training time as we had determined in previous experiments that the main overhead occurred during the first epoch, while timing results for all remaining epochs remained stable. 

\section{Results}
In the section, we discuss initial results generated by our benchmarks. In all figures, the Y-axis denotes the wallclock time in seconds required to complete 12 epochs of \emph{MNIST} training, or the average time per epoch of \emph{ResNet50} training. 
Also \emph{TFx.x} refers to TensorFlow version \emph{x.x}, \emph{src} post-fix refers to the optimal source built container and  \emph{XLA} or \emph{NGRAPH} post-fix denotes is that graph compiler is enabled. PyTorch version 1.4 is used for all comparisons.

\begin{figure}[ht]
    \centering
    \includegraphics[width=9 cm]{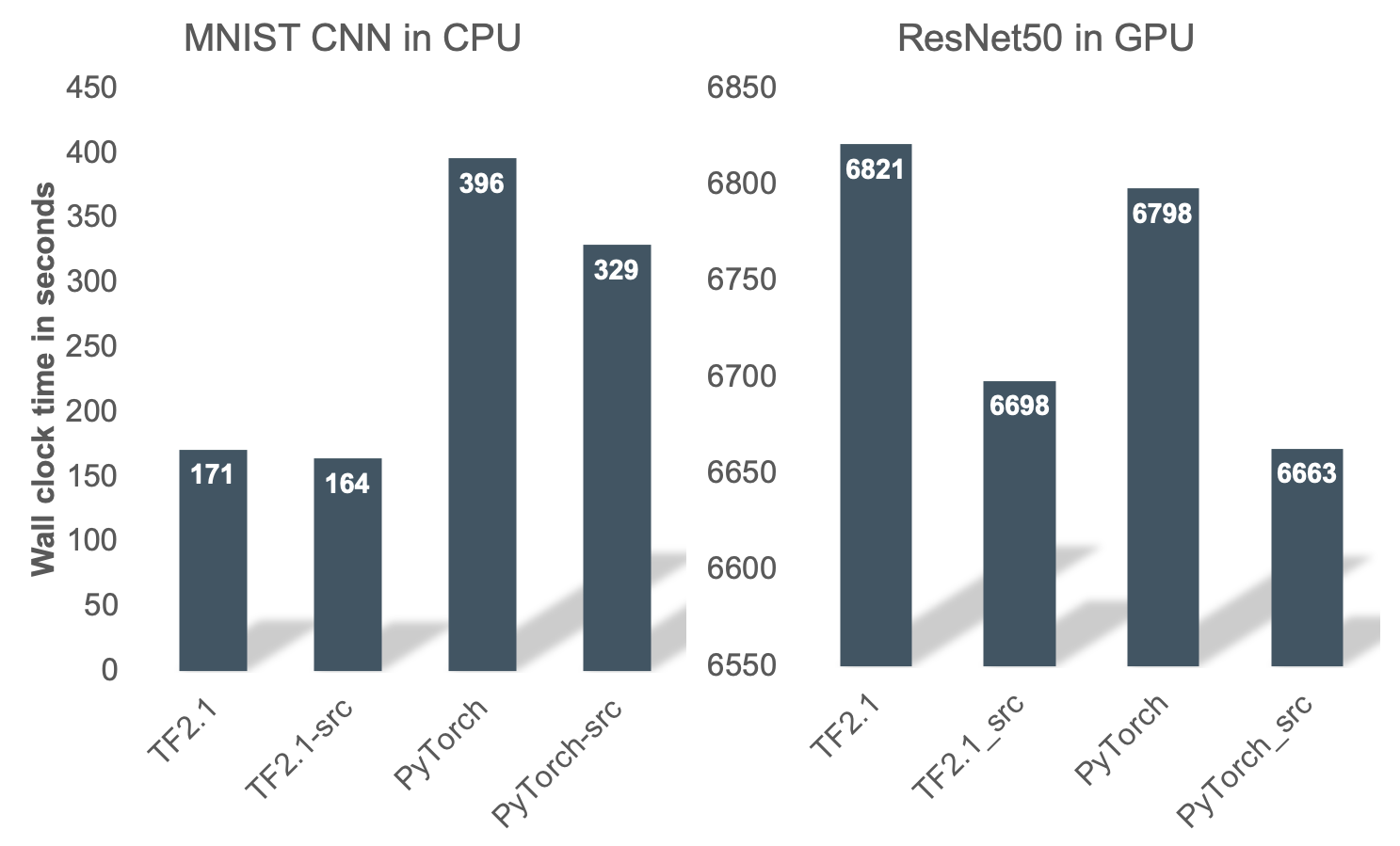}
    \caption{Performance of AI framework containers custom built from source. Left - MNIST CNN training in CPU only. Right- ResNet50 in GPU. }
    \label{fig:mnist_src}
\end{figure}

The results of Figure~\ref{fig:mnist_all} show the performance of the DockerHub containers for \emph{MNIST} CNN training on CPU only. Graph compilers are not enabled for these results. As can be seen, TensorFlow 1.14, PyTorch, and MXNet (v1.6) perform similarly across the board, with the exception of TensorFlow 2.1, which shows a nearly 54\% improvement over TensorFlow 1.14, likely due to eager execution being enabled by default starting from TensorFlow 2.0, while TensorFlow 1.14 uses graph execution. CNTK (v2.7) is a far outlier due to a lack of CPU optimisations, as mentioned in the official documentation. Note that MXNet and CNTK were only evaluated for comparison purposes and no further containers were evaluated beyond those attained from DockerHub.

Figure~\ref{fig:mnist_src} (left) compares the training time of MNIST CNN in custom built containers to that of DockerHub containers. The TensorFlow custom built container shows little improvement (4\%) compared to the DockerHub container, whereas the PyTorch container gives a substantial 17\% speedup over the official DockerHub one. 

Figure~\ref{fig:mnist_src} (right) shows the result of ResNet50 training on \emph{ImageNet} data with a custom built AI framework for the NVIDIA GPUs. 
A slight 2\% performance improvement is visible for both TensorFlow and PyTorch source built containers.  We see a similar performance for MXNet containers.

%\begin{figure}[ht]
%    \centering
%    \includegraphics[width=5 cm]{images/resnet_src.png}
%    \caption{Performance of source built AI framework containers built %on GPU ImageNet workload}
%    \label{fig:resnet_src}
%\end{figure}

\begin{figure}[ht]
    \centering
    \includegraphics[width=7.5 cm]{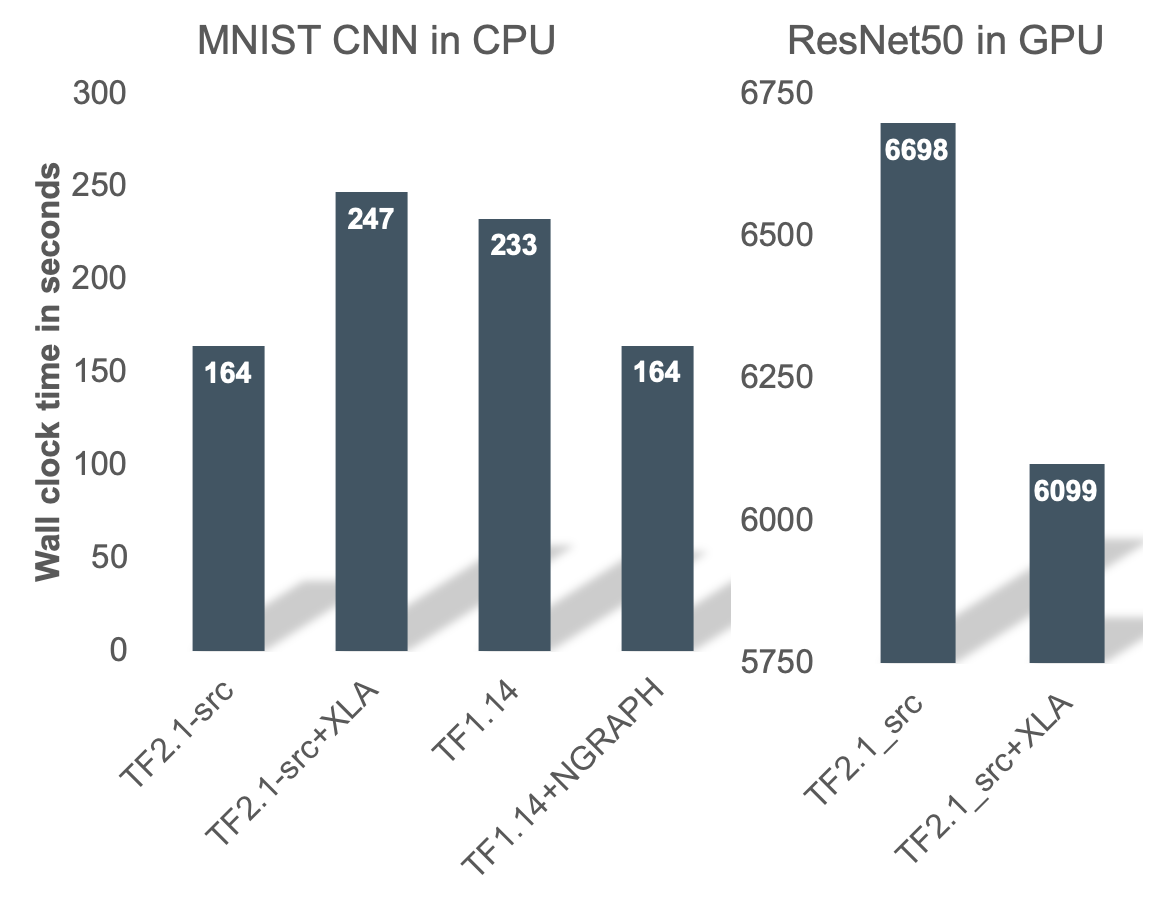}
    \caption{Performance of AI framework containers with graph compilers. Left - MNIST CNN training in CPU only. Right- ResNet50 in GPU. }
    \label{fig:mnist_graph}
\end{figure}

Figure~\ref{fig:mnist_graph} (left) compares the performance of TensorFlow in combination with \emph{XLA} and \emph{nGraph} for MNIST CNN training on a CPU. \emph{XLA} is supported in most versions of TensorFlow, whereas \emph{nGraph} is not yet supported for TensorFlow 2.x.
We evaluated \emph{XLA} on the latest standard release version 2.1, and nGraph on TensorFlow 1.4. A marked performance loss can be observed when running TensorFlow with \emph{XLA} on the CPU. This is likely due to the fact that the \emph{XLA} team has in recent years focused exclusively on the optimisation of just in time compilation on GPUs, and the overhead induced by the additional graph compilations on a simple network like \emph{MNIST}\cite{leary2017xla}. \emph{nGraph}, on the other hand, shows speedup with a 30\% improvement.  

Figure~\ref{fig:mnist_graph} (right) shows the performance of the TensorFlow 2.1 source build using XLA on the ResNet50 for ImageNet workload on a GPU. The performance is improved by 9\% using XLA, which is significantly better than the 30\% performance degradation seen on the CPU.

%\begin{figure}[ht]
%    \centering
%    \includegraphics[width=3 cm]{images/resnet_graph.png}
%    \caption{Performance of AI framework containers with Graph %compilers on GPU ImageNet workload}
%    \label{fig:resnet_graph}
%\end{figure}

We are currently evaluating the GLOW graph compiler container on CPU and GPU, and the \emph{nGraph} graph compiler on the GPU. Thus far, the benchmark results show a trend of custom-built, hardware-targeted containers improving performance. For graph compilers, the optimisation results are dependent on the neural network and the target hardware. 

\section{Conclusions}
Software-defined infrastructures offer flexibility in deploying AI training workloads across heterogeneous targets. With diverse hardware, optimising AI workloads with different configurations and data sets is mandatory. We introduced MODAK, a novel tool that maps optimal application parameters to infrastructure using performance modelling and container technology. We demonstrated the usage of MODAK by optimising AI training deployments with graph compilers and Singularity containers. We showed an up to 17\% speedup using custom built optimised containers and up to a 30\%  speedup using graph compilers. We also found use cases where graph compilers slowed the training. These benchmark results are then used to model the performance and optimise similar training workloads.

\section*{Future Work}
Additional work is required to benchmark \emph{GLOW} and \emph{nGraph} on PyTorch and MXNet, as well as any emerging compilers. We plan to extend our containers to encompass more HPC specific workloads, such as MPI applications. Finally, while Singularity containers are the preferred runtime for HPC, we will port the images to Docker for easier dissemination of the project. 

\section*{Acknowledgment}
This work has received funding from the European Union’s Horizon 2020 research and innovation program under grant agreement No 825480.
We would like to acknowledge colleagues at the HPE HPC/AI EMEA Research lab and SODALITE for their support. We also thank \emph{Tim Dykes} for proofreading this paper and \emph{Irene Ferrario} and \emph{Harvey Richardson} for the support and guidance.

\bibliographystyle{IEEEtran}

\bibliography{IEEEabrv,paper.bib}

\begin{thebibliography}{10}
\providecommand{\url}[1]{#1}
\csname url@samestyle\endcsname
\providecommand{\newblock}{\relax}
\providecommand{\bibinfo}[2]{#2}
\providecommand{\BIBentrySTDinterwordspacing}{\spaceskip=0pt\relax}
\providecommand{\BIBentryALTinterwordstretchfactor}{4}
\providecommand{\BIBentryALTinterwordspacing}{\spaceskip=\fontdimen2\font plus
\BIBentryALTinterwordstretchfactor\fontdimen3\font minus
  \fontdimen4\font\relax}
\providecommand{\BIBforeignlanguage}[2]{{%
\expandafter\ifx\csname l@#1\endcsname\relax
\typeout{** WARNING: IEEEtran.bst: No hyphenation pattern has been}%
\typeout{** loaded for the language `#1'. Using the pattern for}%
\typeout{** the default language instead.}%
\else
\language=\csname l@#1\endcsname
\fi
#2}}
\providecommand{\BIBdecl}{\relax}
\BIBdecl

\bibitem{alom2018history}
M.~Z. Alom, T.~M. Taha, C.~Yakopcic, S.~Westberg, P.~Sidike, M.~S. Nasrin,
  B.~C. Van~Esesn, A.~A.~S. Awwal, and V.~K. Asari, ``The history began from
  alexnet: A comprehensive survey on deep learning approaches,'' \emph{arXiv
  preprint arXiv:1803.01164}, 2018.

\bibitem{lecun2015deep}
Y.~LeCun, Y.~Bengio, and G.~Hinton, ``Deep learning,'' \emph{nature}, vol. 521,
  no. 7553, pp. 436--444, 2015.

\bibitem{fukushima1980neocognitron}
K.~Fukushima, ``Neocognitron: A self-organizing neural network model for a
  mechanism of pattern recognition unaffected by shift in position,''
  \emph{Biological cybernetics}, vol.~36, no.~4, pp. 193--202, 1980.

\bibitem{krizhevsky2012imagenet}
A.~Krizhevsky, I.~Sutskever, and G.~E. Hinton, ``Imagenet classification with
  deep convolutional neural networks,'' in \emph{Advances in neural information
  processing systems}, 2012, pp. 1097--1105.

\bibitem{abadi2016tensorflow}
M.~Abadi, P.~Barham, J.~Chen, Z.~Chen, A.~Davis, J.~Dean, M.~Devin,
  S.~Ghemawat, G.~Irving, M.~Isard \emph{et~al.}, ``Tensorflow: A system for
  large-scale machine learning,'' in \emph{12th $\{$USENIX$\}$ Symposium on
  Operating Systems Design and Implementation ($\{$OSDI$\}$ 16)}, 2016, pp.
  265--283.

\bibitem{NEURIPS2019_9015}
\BIBentryALTinterwordspacing
A.~Paszke, S.~Gross, F.~Massa, A.~Lerer, J.~Bradbury, G.~Chanan, T.~Killeen,
  Z.~Lin, N.~Gimelshein, L.~Antiga, A.~Desmaison, A.~Kopf, E.~Yang, Z.~DeVito,
  M.~Raison, A.~Tejani, S.~Chilamkurthy, B.~Steiner, L.~Fang, J.~Bai, and
  S.~Chintala, ``Pytorch: An imperative style, high-performance deep learning
  library,'' in \emph{Advances in Neural Information Processing Systems 32},
  H.~Wallach, H.~Larochelle, A.~Beygelzimer, F.~d\textquotesingle
  Alch\'{e}-Buc, E.~Fox, and R.~Garnett, Eds.\hskip 1em plus 0.5em minus
  0.4em\relax Curran Associates, Inc., 2019, pp. 8024--8035. [Online].
  Available:
  \url{http://papers.neurips.cc/paper/9015-pytorch-an-imperative-style-high-performance-deep-learning-library.pdf}
\BIBentrySTDinterwordspacing

\bibitem{chen2015mxnet}
T.~Chen, M.~Li, Y.~Li, M.~Lin, N.~Wang, M.~Wang, T.~Xiao, B.~Xu, C.~Zhang, and
  Z.~Zhang, ``Mxnet: A flexible and efficient machine learning library for
  heterogeneous distributed systems,'' \emph{arXiv preprint arXiv:1512.01274},
  2015.

\bibitem{seide2016cntk}
F.~Seide and A.~Agarwal, ``Cntk: Microsoft's open-source deep-learning
  toolkit,'' in \emph{Proceedings of the 22nd ACM SIGKDD International
  Conference on Knowledge Discovery and Data Mining}, 2016, pp. 2135--2135.

\bibitem{chollet2015keras}
F.~Chollet \emph{et~al.}, ``Keras,'' \url{https://keras.io}, 2015.

\bibitem{terraform}
\BIBentryALTinterwordspacing
2020. [Online]. Available: \url{https://www.terraform.io}
\BIBentrySTDinterwordspacing

\bibitem{cloudify_2020}
\BIBentryALTinterwordspacing
2020. [Online]. Available: \url{https://cloudify.co/}
\BIBentrySTDinterwordspacing

\bibitem{Jette02slurm:simple}
M.~A. Jette, A.~B. Yoo, and M.~Grondona, ``Slurm: Simple linux utility for
  resource management,'' in \emph{In Lecture Notes in Computer Science:
  Proceedings of Job Scheduling Strategies for Parallel Processing (JSSPP)
  2003}.\hskip 1em plus 0.5em minus 0.4em\relax Springer-Verlag, 2002, pp.
  44--60.

\bibitem{klusavcek2015planning}
D.~Klus{\'a}{\v{c}}ek, V.~Chlumsk{\`y}, and H.~Rudov{\'a}, ``Planning and
  optimization in torque resource manager,'' in \emph{Proceedings of the 24th
  International Symposium on High-Performance Parallel and Distributed
  Computing}, 2015, pp. 203--206.

\bibitem{hetero}
\BIBentryALTinterwordspacing
2020. [Online]. Available: \url{http://heterogeneityalliance.eu}
\BIBentrySTDinterwordspacing

\bibitem{cola}
\BIBentryALTinterwordspacing
2020. [Online]. Available: \url{https://project-cola.eu}
\BIBentrySTDinterwordspacing

\bibitem{tango}
\BIBentryALTinterwordspacing
2020. [Online]. Available: \url{http://www.tango-project.eu}
\BIBentrySTDinterwordspacing

\bibitem{hidalgo}
\BIBentryALTinterwordspacing
2020. [Online]. Available: \url{https://hidalgo-project.eu}
\BIBentrySTDinterwordspacing

\bibitem{exa2pro}
\BIBentryALTinterwordspacing
2020. [Online]. Available: \url{https://exa2pro.eu}
\BIBentrySTDinterwordspacing

\bibitem{ecoscale}
\BIBentryALTinterwordspacing
2020. [Online]. Available: \url{www.ecoscale.eu}
\BIBentrySTDinterwordspacing

\bibitem{sodalitewebsite}
Sodalite. \url{https://www.sodalite.eu/}, last accessed 10 June 2020.

\bibitem{baughman2018profiling}
M.~Baughman, R.~Chard, L.~T. Ward, J.~Pitt, K.~Chard, and I.~T. Foster,
  ``Profiling and predicting application performance on the cloud.'' in
  \emph{UCC}, 2018, pp. 21--30.

\bibitem{wu2019paraopt}
C.~Wu, T.~Summer, Z.~Li, A.~Woodard, R.~Chard, M.~Baughman, Y.~Babuji,
  K.~Chard, J.~Pitt, and I.~Foster, ``Paraopt: Automated application
  parameterization and optimization for the cloud,'' in \emph{2019 IEEE
  International Conference on Cloud Computing Technology and Science
  (CloudCom)}.\hskip 1em plus 0.5em minus 0.4em\relax IEEE, 2019, pp. 255--262.

\bibitem{mohammadi2018comparative}
M.~Mohammadi and T.~Bazhirov, ``Comparative benchmarking of cloud computing
  vendors with high performance linpack,'' in \emph{Proceedings of the 2nd
  International Conference on High Performance Compilation, Computing and
  Communications}, 2018, pp. 1--5.

\bibitem{mohammadi2018continuous}
------, ``Continuous evaluation of the performance of cloud infrastructure for
  scientific applications,'' \emph{arXiv preprint arXiv:1812.05257}, 2018.

\bibitem{epcc}
\BIBentryALTinterwordspacing
2020. [Online]. Available:
  \url{https://www.epcc.ed.ac.uk/blog/2020/06/benchmarking-oracle-bare-metal-cloud-dirac-hpc-workloads}
\BIBentrySTDinterwordspacing

\bibitem{chiba2019confadvisor}
T.~Chiba, R.~Nakazawa, H.~Horii, S.~Suneja, and S.~Seelam, ``Confadvisor: A
  performance-centric configuration tuning framework for containers on
  kubernetes,'' in \emph{2019 IEEE International Conference on Cloud
  Engineering (IC2E)}.\hskip 1em plus 0.5em minus 0.4em\relax IEEE, 2019, pp.
  168--178.

\bibitem{awscompute}
Aws compute optimizer. \url{https://aws.amazon.com/compute-optimizer/}, last
  accessed 12. June 2020.

\bibitem{krishnan2015google}
S.~Krishnan and J.~L.~U. Gonzalez, ``Google compute engine,'' in \emph{Building
  your next big thing with Google cloud platform}.\hskip 1em plus 0.5em minus
  0.4em\relax Springer, 2015, pp. 53--81.

\bibitem{brayford2019deploying}
D.~Brayford, S.~Vallecorsa, A.~Atanasov, F.~Baruffa, and W.~Riviera,
  ``Deploying ai frameworks on secure hpc systems with containers.'' in
  \emph{2019 IEEE High Performance Extreme Computing Conference (HPEC)}.\hskip
  1em plus 0.5em minus 0.4em\relax IEEE, 2019, pp. 1--6.

\bibitem{priedhorsky2017charliecloud}
R.~Priedhorsky and T.~Randles, ``Charliecloud: Unprivileged containers for
  user-defined software stacks in hpc,'' in \emph{Proceedings of the
  International Conference for High Performance Computing, Networking, Storage
  and Analysis}, 2017, pp. 1--10.

\bibitem{modak}
\BIBentryALTinterwordspacing
K.~Sivalingam \emph{et~al.}, ``Prototype of application and infrastructure
  performance models,'' in \emph{SODALITE Deliverables}.\hskip 1em plus 0.5em
  minus 0.4em\relax EC, 2020. [Online]. Available:
  \url{https://www.sodalite.eu/deliverables}
\BIBentrySTDinterwordspacing

\bibitem{bernstein2014containers}
D.~Bernstein, ``Containers and cloud: From lxc to docker to kubernetes,''
  \emph{IEEE Cloud Computing}, vol.~1, no.~3, pp. 81--84, 2014.

\bibitem{helsley2009lxc}
M.~Helsley, ``Lxc: Linux container tools,'' \emph{IBM devloperWorks Technical
  Library}, vol.~11, 2009.

\bibitem{joy2015performance}
A.~M. Joy, ``Performance comparison between linux containers and virtual
  machines,'' in \emph{2015 International Conference on Advances in Computer
  Engineering and Applications}.\hskip 1em plus 0.5em minus 0.4em\relax IEEE,
  2015, pp. 342--346.

\bibitem{merkel2014docker}
D.~Merkel, ``Docker: lightweight linux containers for consistent development
  and deployment,'' \emph{Linux journal}, vol. 2014, no. 239, p.~2, 2014.

\bibitem{gerhardt2017shifter}
L.~Gerhardt, W.~Bhimji, M.~Fasel, J.~Porter, M.~Mustafa, D.~Jacobsen,
  V.~Tsulaia, and S.~Canon, ``Shifter: Containers for hpc,'' in \emph{J. Phys.
  Conf. Ser.}, vol. 898, 2017, p. 082021.

\bibitem{sylabswebsite}
Singularity. \url{https://sylabs.io/singularity/}, last accessed 10. June 2020.

\bibitem{kurtzer2017singularity}
G.~M. Kurtzer, V.~Sochat, and M.~W. Bauer, ``Singularity: Scientific containers
  for mobility of compute,'' \emph{PloS one}, vol.~12, no.~5, 2017.

\bibitem{rudyy2019containers}
O.~Rudyy, M.~Garcia-Gasulla, F.~Mantovani, A.~Santiago, R.~Sirvent, and
  M.~V{\'a}zquez, ``Containers in hpc: A scalability and portability study in
  production biological simulations,'' in \emph{2019 IEEE International
  Parallel and Distributed Processing Symposium (IPDPS)}.\hskip 1em plus 0.5em
  minus 0.4em\relax IEEE, 2019, pp. 567--577.

\bibitem{benedicic2019sarus}
L.~Benedicic, F.~A. Cruz, A.~Madonna, and K.~Mariotti, ``Sarus: Highly scalable
  docker containers for hpc systems,'' in \emph{International Conference on
  High Performance Computing}.\hskip 1em plus 0.5em minus 0.4em\relax Springer,
  2019, pp. 46--60.

\bibitem{TFXLAwebsite}
Xla: Optimizing compiler for machine learning : Tensorflow.
  \url{https://www.tensorflow.org/xla}, last accessed 15. June 2020.

\bibitem{leary2017xla}
C.~Leary and T.~Wang, ``Xla: Tensorflow, compiled,'' \emph{TensorFlow Dev
  Summit}, 2017.

\bibitem{rotem2018glow}
N.~Rotem, J.~Fix, S.~Abdulrasool, G.~Catron, S.~Deng, R.~Dzhabarov, N.~Gibson,
  J.~Hegeman, M.~Lele, R.~Levenstein \emph{et~al.}, ``Glow: Graph lowering
  compiler techniques for neural networks,'' \emph{arXiv preprint
  arXiv:1805.00907}, 2018.

\bibitem{cyphers2018intel}
S.~Cyphers, A.~K. Bansal, A.~Bhiwandiwalla, J.~Bobba, M.~Brookhart,
  A.~Chakraborty, W.~Constable, C.~Convey, L.~Cook, O.~Kanawi \emph{et~al.},
  ``Intel ngraph: An intermediate representation, compiler, and executor for
  deep learning,'' \emph{arXiv preprint arXiv:1801.08058}, 2018.

\bibitem{lin2019onnc}
W.-F. Lin, D.-Y. Tsai, L.~Tang, C.-T. Hsieh, C.-Y. Chou, P.-H. Chang, and
  L.~Hsu, ``Onnc: A compilation framework connecting onnx to proprietary deep
  learning accelerators,'' in \emph{2019 IEEE International Conference on
  Artificial Intelligence Circuits and Systems (AICAS)}.\hskip 1em plus 0.5em
  minus 0.4em\relax IEEE, 2019, pp. 214--218.

\bibitem{wang2014intel}
E.~Wang, Q.~Zhang, B.~Shen, G.~Zhang, X.~Lu, Q.~Wu, and Y.~Wang, ``Intel math
  kernel library,'' in \emph{High-Performance Computing on the
  Intel{\textregistered} Xeon Phi™}.\hskip 1em plus 0.5em minus 0.4em\relax
  Springer, 2014, pp. 167--188.

\bibitem{brown2015gpu}
L.~Brown, ``Gpu accelerated deep learning with cudnn,'' \emph{GTC}, 2015.

\bibitem{hlrs}
Hlrs - high-performance computing center | stuttgart.
  \url{https://www.hlrs.de/home/}, last accessed 17. June 2020.

\bibitem{singularitymapping}
User namespaces and fakeroot.
  \url{https://sylabs.io/guides/3.5/admin-guide/user\_namespace.html\#adding-a-fakeroot-mapping},
  last accessed 15. June 2020.

\bibitem{tfsourcebuild}
Tensorflow: Build from source. \url{https://www.tensorflow.org/install/source},
  last accessed 10. June 2020.

\bibitem{torchsourcebuild}
Pytorch: From source. \url{https://github.com/pytorch/pytorch\#from-source},
  last accessed 10. June 2020.

\bibitem{glowsourcebuild}
Glow. \url{https://github.com/pytorch/glow}, last accessed 10. June 2020.

\bibitem{lecun-mnisthandwrittendigit-2010}
\BIBentryALTinterwordspacing
Y.~LeCun and C.~Cortes, ``{MNIST} handwritten digit database,'' 2010. [Online].
  Available: \url{http://yann.lecun.com/exdb/mnist/}
\BIBentrySTDinterwordspacing

\bibitem{deng2009imagenet}
J.~Deng, W.~Dong, R.~Socher, L.-J. Li, K.~Li, and L.~Fei-Fei, ``Imagenet: A
  large-scale hierarchical image database,'' in \emph{2009 IEEE conference on
  computer vision and pattern recognition}.\hskip 1em plus 0.5em minus
  0.4em\relax Ieee, 2009, pp. 248--255.

\bibitem{deng2012mnist}
L.~Deng, ``The mnist database of handwritten digit images for machine learning
  research [best of the web],'' \emph{IEEE Signal Processing Magazine},
  vol.~29, no.~6, pp. 141--142, 2012.

\bibitem{lecun1998gradient}
Y.~LeCun, L.~Bottou, Y.~Bengio, and P.~Haffner, ``Gradient-based learning
  applied to document recognition,'' \emph{Proceedings of the IEEE}, vol.~86,
  no.~11, pp. 2278--2324, 1998.

\bibitem{russakovsky2015imagenet}
O.~Russakovsky, J.~Deng, H.~Su, J.~Krause, S.~Satheesh, S.~Ma, Z.~Huang,
  A.~Karpathy, A.~Khosla, M.~Bernstein \emph{et~al.}, ``Imagenet large scale
  visual recognition challenge,'' \emph{International journal of computer
  vision}, vol. 115, no.~3, pp. 211--252, 2015.

\bibitem{he2016deep}
K.~He, X.~Zhang, S.~Ren, and J.~Sun, ``Deep residual learning for image
  recognition,'' in \emph{Proceedings of the IEEE conference on computer vision
  and pattern recognition}, 2016, pp. 770--778.

\bibitem{hochreiter2001gradient}
S.~Hochreiter, Y.~Bengio, P.~Frasconi, J.~Schmidhuber \emph{et~al.}, ``Gradient
  flow in recurrent nets: the difficulty of learning long-term dependencies,''
  2001.

\end{thebibliography}

\vspace{12pt}

\end{document}